\documentclass[a4paper,10pt]{article}
\usepackage{stmaryrd}
\usepackage{amsfonts}
\usepackage{bbm}
\usepackage{amscd}
\usepackage{mathrsfs}
\usepackage{latexsym,amssymb,amsmath,amscd,amscd,amsthm,amsxtra}
\usepackage[dvips]{graphicx}
\usepackage[utf8]{inputenc}
\usepackage[T1]{fontenc}
\usepackage{lmodern}
\usepackage{amssymb}
\usepackage[all]{xy}
\usepackage{nicefrac,mathtools,enumitem}
\usepackage{microtype}
\usepackage{xcolor}
\newtheorem{thm}{Theorem}[section]
\newtheorem{lem}[thm]{Lemma}
\newtheorem{cor}[thm]{Corollary}
\newtheorem{pro}[thm]{Proposition}
\newtheorem{ex}[thm]{Example}
\newtheorem{rmk}[thm]{Remark}
\newtheorem{defi}[thm]{Definition}

\newcommand {\comment}[1]{{\marginpar{*}\scriptsize\textbf{Comments:} #1}}

\newcommand {\emptycomment}[1]{} 

\newcommand{\be }{\begin{equation}}
\newcommand{\ee }{\end{equation}}

\newcommand{\pf}{\noindent{\bf Proof.}\ }

\newcommand{\Sym}{\mathsf{Sym}}

\newcommand{\huaB}{\mathcal{B}}

\newcommand{\huaA}{\mathcal{A}}
\newcommand{\huaL}{\mathcal{L}}

\newcommand{\huaF}{\mathcal{F}}

\newcommand{\CWM}{C^{\infty}(M)}

\newcommand{\frkg}{\mathfrak g}

\newcommand{\frkD}{\mathfrak D}

\newcommand{\frkL}{\mathfrak L}

\def\qed{\hfill ~\vrule height6pt width6pt depth0pt}

\newcommand{\half}{\frac{1}{2}}

\newcommand{\br}[1]{   [ \cdot,    \cdot  ]   }

\newcommand{\dev}{\mathfrak{D}}

\newcommand{\id}{\rm{id}}

\newcommand{\g}{\mathfrak g}

\newcommand{\dM}{\mathrm{d}}

\newcommand{\Hom}{\mathrm{Hom}}

\newcommand{\End}{\mathrm{End}}

\begin{document}
\title{Left-symmetric bialgebroids and their corresponding Manin
triples
\thanks
 {
This research  is supported by
NSF of China (11471139, 11271202, 11221091, 11425104), SRFDP
(20120031110022) and NSF of Jilin Province (20140520054JH).
 }}
\author{Jiefeng Liu$^1$, Yunhe Sheng$^{1,2}$ and Chengming Bai$^3$\\
$^1$Department of Mathematics, Xinyang Normal University,\\ \vspace{2mm} Xinyang 464000, Henan, China\\
 $^2$Department of Mathematics, Jilin University,
 \\\vspace{2mm}Changchun 130012, Jilin, China\\
$^3$Chern Institute of Mathematics and LPMC, Nankai University,\\
Tianjin 300071, China \\ Email:  jfliu12@126.com; shengyh@jlu.edu.cn; baicm@nankai.edu.cn \\
}

\date{}
\footnotetext{{\it{Keywords}: left-symmetric algebroid,  left-symmetric bialgebroid, pseudo-Hessian manifold,  pre-symplectic algebroid, Manin triple   }}
\footnotetext{{\it{MSC}}: 17B62,53D12,53D17,53D18}

\maketitle
\begin{abstract}
 In this paper, we introduce the notion of a left-symmetric
bialgebroid as a geometric generalization of a left-symmetric
bialgebra and construct a left-symmetric bialgebroid from a
pseudo-Hessian manifold. We also introduce the notion of a
Manin triple for left-symmetric algebroids, which is equivalent to
a left-symmetric bialgebroid.  The corresponding double structure
 is a pre-symplectic algebroid
rather than a left-symmetric algebroid. In particular,  we
establish a relation between  Maurer-Cartan type equations and
Dirac structures of the pre-symplectic algebroid which is the
corresponding double structure for a left-symmetric bialgebroid.
\end{abstract}

\section{Introduction}

  Left-symmetric algebras (or pre-Lie algebras) arose from the
study of convex homogeneous cones \cite{Vinb}, affine manifolds
and affine structures on Lie groups \cite{Koszul1}, deformation
and cohomology theory of associative algebras \cite{G} and then
appear in many fields in mathematics and mathematical physics. See
the survey article \cite{Pre-lie algebra in geometry} and the
references therein. In particular, there are
 close relations between left-symmetric algebras and
certain important left-invariant structures on Lie groups like
aforementioned affine, symplectic, K\"ahler, and metric structures
\cite{Kim,Lichnerowicz,MT,Med,Milnor}. A quadratic left-symmetric algebra is a left-symmetric algebra together with a nondegenerate invariant skew-symmetric bilinear form \cite{symplectic Lie algebras}. A symplectic (Frobenius) Lie algebra  is a Lie algebra $\g$ equipped with a nondegenerate 2-cocycle $\omega\in\wedge^2\g^*$.
There is a one-to-one correspondence between symplectic (Frobenius) Lie algebras and  quadratic left-symmetric algebras.

A
left-symmetric algebroid, also called a Koszul-Vinberg algebroid, is a geometric generalization of a
left-symmetric algebra. See \cite{LiuShengBaiChen,Boyom1,Boyom2} for more details and applications.
In \cite{lsb}, we introduced the notion of a pre-symplectic algebroid,
which is a geometric generalization of a quadratic left-symmetric
algebra. Generalizing the relation between symplectic~ (Frobenius)~ Lie~
algebras and quadratic  left-symmetric  algebras, we showed that there is a one-to-one correspondence between
symplectic Lie algebroids and pre-symplectic algebroids.
See \cite{symplectic lie algebroid,reduction,reduction1,NestTsygan} for more details about symplectic Lie algebroids and their applications.

The purpose of this paper is studying the bialgebra theory for left-symmetric algebroids and the corresponding Manin triple theory. Motivated by \cite{lwx,Lie bialgebroid}, we introduce the notion of a left-symmetric bialgebroid, which is a geometric generalization of a left-symmetric bialgebra \cite{Left-symmetric bialgebras}.
The double of a left-symmetric bialgebroid is not a left-symmetric
algebroid anymore, but a pre-symplectic algebroid. This result is  parallel to the fact that
 the double of a Lie bialgebroid\footnote{The notion of a Lie bialgebroid was first introduced by Mackenzie and Xu in as the infinitesimal object of a Poisson groupoid \cite{Lie bialgebroid}.}, is not a Lie algebroid, but a Courant algebroid \cite{lwx}.
Furthermore, if we consider the commutator of a left-symmetric
bialgebroid, we can obtain a matched pair of Lie algebroids, whose
double is the symplectic Lie algebroid associated to the
pre-symplectic algebroid. The above results can be summarized into
the following   commutative  diagram:

{\footnotesize{
$$
\xymatrix{
&\mbox{qua LS alg~}\g\oplus\g^*\ar[rr]^{\mbox{commutator}}\ar[dd]&&\mbox{sym Lie  alg~}\g^c\oplus {\g^*}^c\ar[dd]\\
\mbox{LS bialg~}(\g,\g^*)\ar[dd]\ar[rr]^{\mbox{commutator}}\ar[ur]^{\mbox{double}}&&\mbox{MP~ Lie alg~}(\g^c,{\g^*}^c)\ar[dd]\ar[ur]^{\mbox{double}}&\\
&\mbox{pre-sym algd}~A\oplus A^*\ar[rr]^{\mbox{commutator}}&&\mbox{sym Lie algd}~A^c\oplus {A^*}^c\\
\mbox{LS bialgd~}(A,A^*)\ar[rr]^{\mbox{commutator}}\ar[ur]^{\mbox{double}}&&\mbox{MP ~Lie algd~}(A^c, {A^*}^c)\ar[ur]^{\mbox{double}}&
}
$$
}}
$$
 \footnotesize{ \mbox{ Diagram~ I}}
$$
In the above diagram, qua is short for quadratic, LS is short for left-symmetric, alg is short for  algebra, sym is short for symplectic,  MP is short for matched pair and algd is short for algebroid. We establish a relation between left-symmetric bialgebroids and pseudo-Hessian manifolds. See \cite{NiBai,Shima,Geometry of Hessian structures} for more information about pseudo-Hessian Lie algebras and Hessian geometry.
A flat manifold $(M,\nabla)$ gives rise to a left-symmetric algebroid $T_\nabla M$.  We show that a pseudo-Riemannian metric $g$ on $(M,\nabla)$ is a pseudo-Hessian metric if $\delta g=0$, where $\delta$ is the cohomology operator of the left-symmetric algebroid $T_\nabla M$ (Proposition \ref{pro:hessian-cocycle}). 
 Given a pseudo-Hessian manifold $(M,\nabla,g)$, $(T_\nabla M,T^*_HM)$ is a left-symmetric bialgebroid (Proposition \ref{cor:hessianLA}), where $H$ is the inverse of $g$. This result is parallel to
 that $(TM, T^*_\pi M)$ is a Lie bialgebroid for any Poisson manifold $(M,\pi)$ \cite{KS,Lie bialgebroid}. It seems that our theory is a symmetric analogue of Poisson geometry.


The paper is organized as follows. In Section $2$, we give a review on  Lie algebroids, left-symmetric algebroids and pre-symplectic algebroids. In Section $3$, we develop the differential calculus on a left-symmetric algebroid which is the main tool in our later study. In Section $4$,  we introduce the notion of a left-symmetric bialgebroid and study its properties. In Section 5, we introduce the notion of a Manin triple for left-symmetric algebroids and show the equivalence between  left-symmetric bialgebroids and Manin triples for left-symmetric algebroids.

Throughout this paper,  all the vector bundles are over the same manifold
$M$. For two vector bundles $A$ and $B$, a bundle map from $A$ to
$B$ is a base-preserving map and $\CWM$-linear.

\vspace{2mm}
 \noindent {\bf Acknowledgement:} We give our warmest thanks to Zhangju Liu and Jianghua Lu for very useful comments and discussions.

\section{Preliminaries}

We briefly recall Lie algebroids, left-symmetric algebroids and pre-symplectic algebroids.

\subsection*{Lie algebroids}

The notion of a Lie algebroid was introduced
by Pradines in 1967, which is a generalization of Lie algebras and
tangent bundles. See \cite{General theory of Lie groupoid and Lie
algebroid} for general theory about Lie algebroids. They play
important roles in various parts of mathematics.
\begin{defi}
A {\bf Lie algebroid} structure on a vector bundle $\huaA\longrightarrow M$ is
a pair that consists of a Lie algebra structure $[\cdot,\cdot]_\huaA$ on
the section space $\Gamma(\huaA)$ and a  bundle map
$a_\huaA:\huaA\longrightarrow TM$, called the anchor, such that the
following relation is satisfied:
$$~[x,fy]_\huaA=f[x,y]_\huaA+a_\huaA(x)(f)y,\quad \forall~f\in
\CWM.$$
\end{defi}

For a vector bundle $E\longrightarrow M$, we denote by
$\dev(E)$   the gauge Lie algebroid of the
 frame bundle
 $\huaF(E)$, which is also called the covariant differential operator bundle of $E$.

Let $(\huaA,[\cdot,\cdot]_\huaA,a_\huaA)$ and $(\huaB,[\cdot,\cdot]_\huaB,a_\huaB)$ be two Lie
algebroids (with the same base), a {\bf base-preserving morphism}
from $\huaA$ to $\huaB$ is a bundle map $\sigma:\huaA\longrightarrow \huaB$ such
that
\begin{eqnarray*}
  a_\huaB\circ\sigma=a_\huaA,\quad
  \sigma[x,y]_\huaA=[\sigma(x),\sigma(y)]_\huaB.
\end{eqnarray*}

A {\bf representation} of a Lie algebroid $\huaA$
 on a vector bundle $E$ is a base-preserving morphism $\rho$ form $\huaA$ to the Lie algebroid $\dev(E)$.
Denote a representation by $(E;\rho).$
The {\bf dual representation} of a Lie algebroid $\huaA$ on $E^*$ is the bundle map $\rho^*:\huaA\longrightarrow \dev(E^*)$ given by
$$
\langle \rho^*(x)(\xi),u\rangle=a_\huaA(x)\langle \xi,u\rangle-\langle \xi,\rho(x)(u)\rangle,\quad \forall~x\in \Gamma(\huaA),~\xi\in\Gamma(E^*),u\in\Gamma(E).
$$

As a generalization of a matched pair of Lie algebras, a {\bf matched pair of Lie algebroids} is a pair of Lie algebroids $(\huaA,\huaB)$ together with two representations $\rho_\huaA:\huaA\longrightarrow\dev(\huaB)$ and $\rho_\huaB:\huaB\longrightarrow\dev(\huaA)$ such that some compatibility conditions are satisfied.

For all $x\in \Gamma(\huaA)$, the {\bf Lie derivation} $\huaL_x:\Gamma(\huaA^*)\longrightarrow\Gamma(\huaA^*)$ of the Lie algebroid $\huaA$  is given by
\begin{eqnarray}\label{Lie der1}
\langle\huaL_x\xi,y\rangle=a_\huaA(x)\langle \xi,y\rangle-\langle \xi,[x,y]_\huaA\rangle,\quad \forall y\in\Gamma(\huaA),\xi\in\Gamma(\huaA^*).
\end{eqnarray}

A Lie algebroid $(\huaA,[\cdot,\cdot]_\huaA,a_\huaA)$ naturally
represents on the trivial line bundle $E=M\times \mathbb R$ via
the anchor map $a_\huaA:\huaA\longrightarrow TM$. The
corresponding coboundary operator
$\dM:\Gamma(\wedge^k\huaA^*)\longrightarrow
\Gamma(\wedge^{k+1}\huaA^*)$ is given by
\begin{eqnarray*}
  \dM\varpi(x_1,\cdots,x_{k+1})&=&\sum_{i=1}^{k+1}(-1)^{i+1}a_\huaA(x_i)\varpi(x_1\cdots,\widehat{x_i},\cdots,x_{k+1})\\
  &&+\sum_{i<j}(-1)^{i+j}\varpi([x_i,x_j]_\huaA,x_1\cdots,\widehat{x_i},\cdots,\widehat{x_j},\cdots,x_{k+1}).
\end{eqnarray*}
In particular, a $2$-form $\varpi\in\Gamma(\wedge^2\huaA^*)$ is a {\bf 2-cocycle} if $\dM \varpi=0$, i.e.
\begin{equation}
  a_\huaA(x)\varpi(y,z)- a_\huaA(y)\varpi(x,z)+ a_\huaA(z)\varpi(x,y)-\varpi([x,y]_\huaA,z)+\varpi([x,z]_\huaA,y)-\varpi([y,z]_\huaA,x)=0.
\end{equation}

A {\bf symplectic Lie algebroid} is a Lie algebroid together with
a nondegenerate closed $2$-form. A subalgebroid of a symplectic
Lie algebroid $(\huaA,[\cdot,\cdot]_\huaA,a_\huaA,\varpi)$ is
called {\bf Lagrangian} if it is maximal isotropic with respect to the skew-symmetric
bilinear form $\varpi$.

\subsection*{Left-symmetric algebroids}

\begin{defi}
A {\bf left-symmetric algebra} is a pair $(\frkg,\cdot_\frkg)$, where $\g$ is a vector space, and  $\cdot_\frkg:\g\otimes \g\longrightarrow\g$ is a bilinear multiplication
satisfying that for all $x,y,z\in \g$, the associator
\begin{equation}\label{eq:associator}
(x,y,z)\triangleq x\cdot_\frkg(y\cdot_\frkg z)-(x\cdot_\frkg y)\cdot_\frkg z
\end{equation} is symmetric in $x,y$,
i.e.
$$(x,y,z)=(y,x,z),\;\;{\rm or}\;\;{\rm
equivalently,}\;\;x\cdot_\frkg(y\cdot_\frkg z)-(x\cdot_\frkg y)\cdot_\frkg z=y\cdot_\frkg(x\cdot_\frkg z)-(y\cdot_\frkg x)\cdot_\frkg
z.$$
\end{defi}

A left-symmetric algebroid is also called a Koszul-Vinberg algebroid in \cite{Boyom1}.

\begin{defi}{\rm\cite{LiuShengBaiChen,Boyom1}}\label{defi:left-symmetric algebroid}
A {\bf left-symmetric algebroid} structure on a vector bundle
$A\longrightarrow M$ is a pair that consists of a left-symmetric
algebra structure $\cdot_A$ on the section space $\Gamma(A)$ and a
vector bundle morphism $a_A:A\longrightarrow TM$, called the anchor,
such that for all $f\in\CWM$ and $x,y\in\Gamma(A)$, the following
conditions are satisfied:
\begin{itemize}
\item[\rm(i)]$~x\cdot_A(fy)=f(x\cdot_A y)+a_A(x)(f)y,$
\item[\rm(ii)] $(fx)\cdot_A y=f(x\cdot_A y).$
\end{itemize}
\end{defi}

We usually denote a left-symmetric algebroid by $(A,\cdot_A, a_A)$.
Any left-symmetric  algebra is a left-symmetric algebroid over a point.

 \begin{ex}\label{ex:main}{\rm Let $M$
be a differential manifold with a flat torsion free connection
$\nabla$. Then $(TM,\nabla,\id)$ is a left-symmetric algebroid whose sub-adjacent Lie algebroid is
exactly the tangent Lie algebroid. We denote this left-symmetric algebroid by $T_\nabla M$, which will be frequently used below.
}
\end{ex}

For any $x\in\Gamma(A)$, we define
$L_x:\Gamma(A)\longrightarrow\Gamma(A)$  and $R_x:\Gamma(A)\longrightarrow\Gamma(A)$ by
\begin{equation}\label{eq:leftmul}L_xy=x\cdot_A y,\quad R_xy=y\cdot_A x.\end{equation}
Condition (i) in the above definition means that $L_x\in \frkD(A)$.
Condition (ii) means that the map $x\longmapsto L_x$ is
$C^\infty(M)$-linear. Thus, $L:A\longrightarrow \frkD(A)$ is a bundle
map.
\begin{pro}{\rm \cite{LiuShengBaiChen}}\label{thm:sub-adjacent}
  Let $(A,\cdot_A, a_A)$ be a left-symmetric algebroid. Define  a skew-symmetric bilinear bracket operation $[\cdot,\cdot]_A$ on $\Gamma(A)$ by
  $$
  [x,y]_A=x\cdot_A y-y\cdot_A x,\quad \forall ~x,y\in\Gamma(A).
  $$
Then, $(A,[\cdot,\cdot]_A,a_A)$ is a Lie algebroid, and denoted by
$A^c$, called the {\bf sub-adjacent Lie algebroid} of
 $(A,\cdot_A,a_A)$. Furthermore, $L:A\longrightarrow \frkD(A)$  gives a
  representation of the Lie algebroid  $A^c$.
\end{pro}

\begin{thm}{\rm \cite{LiuShengBaiChen}}\label{thm:symLie}
   Let $(A,\cdot_A, a_A)$ be a left-symmetric algebroid. Then $(A^c\ltimes_{L^*}A^*,[\cdot,\cdot]_S,\rho,\omega)$ is a symplectic Lie algebroid, where $A^c\ltimes_{L^*}A^*$ is the semidirect product of $A^c$ and $A^*$ in which $L^*$ is the dual representation of $L$. More precisely, the Lie bracket $[\cdot,\cdot]_S$ and the anchor $\rho$ are given by
   $$
   [x+\xi,y+\eta]_S=[x,y]_A+L^*_x\eta-L^*_y\xi,
   $$
 and $\rho(x+\xi)=a_A(x)$  respectively.
   Furthermore, the symplectic form $\omega$ is given by
   \begin{equation}\label{eq:defiomega}
     \omega(x+\xi,y+\eta)=\langle\xi,y\rangle-\langle\eta,x\rangle,\quad \forall x,y\in\Gamma(A),~\xi,\eta\in\Gamma(A^*).
   \end{equation}
\end{thm}



Let $(A,\cdot_A,a_A)$ be a left-symmetric algebroid and $E$  a vector
bundle. A {\bf representation} of $A$ on $E$ consists of a pair
$(\rho,\mu)$, where $\rho:A\longrightarrow \frkD(E)$ is a representation
of $A^c$ on $E $ and $\mu:A\longrightarrow \End(E)$ is a bundle
map, such that for all $x,y\in \Gamma(A),\ e\in\Gamma(E)$, we have
\begin{eqnarray}\label{representation condition 2}
 \rho(x)\mu(y)e-\mu(y)\rho(x)e=\mu(x\cdot_A y)e-\mu(y)\mu(x)e.
\end{eqnarray}
Denote a representation by $(E;\rho,\mu)$.

\emptycomment{let us recall the cohomology complex for a left-symmetric algebroid $(A,\cdot_A,a_A)$ with a representation $(E;\rho,\mu)$ briefly. Denote the set of $(n+1)$-cochains by
$$C^{n+1}(A,E)=\Gamma(\Hom(\wedge^{n}A\otimes A,E)),\
n\geq 0.$$  For all $\varphi\in C^{n}(A,E)$, the coboundary operator $\delta:C^{n}(A,E)\longrightarrow C^{n+1}(A,E)$ is given by
 \begin{eqnarray*}
\delta\varphi(x_1,\cdots,x_{n+1})&=&\sum_{i=1}^{n}(-1)^{i+1}\rho(x_i)\omega(x_1,\cdots,\hat{x_i},\cdots,x_{n+1})\\
 &&+\sum_{i=1}^{n}(-1)^{i+1}\mu(x_{n+1})\omega(x_1,\cdots,\hat{x_i},\cdots,x_n,x_i)\\
 &&-\sum_{i=1}^{n}(-1)^{i+1}\varphi(x_1,\cdots,\hat{x_i},\cdots,x_n,x_i\cdot_A x_{n+1})\\
 &&+\sum_{1\leq i<j\leq n}(-1)^{i+j}\varphi([x_i,x_j]_A,x_1,\cdots,\hat{x_i},\cdots,\hat{x_j},\cdots,x_{n+1}),
\end{eqnarray*}
for all $x_i\in \Gamma(A),i=1,\cdots,n+1$.
}

Let us recall the cohomology complex with the coefficients in the trivial representation, i.e. $\rho=a_A$ and $\mu=0$. See \cite{cohomology of pre-Lie,LiuShengBaiChen} for general theory of cohomologies of right-symmetric algebras and left-symmetric algebroids respectively. The set of $(n+1)$-cochains  is given by
$$C^{n+1}(A)=\Gamma(\wedge^{n}A^*\otimes A^*),\
n\geq 0.$$  For all $\varphi\in C^{n}(A)$ and $x_i\in
\Gamma(A),~i=1,\cdots,n+1$, the corresponding coboundary operator
$\delta$ is given by
 \begin{eqnarray}\label{LSCA cohomology}
\nonumber\delta\varphi(x_1, \cdots,x_{n+1})
 &=&\sum_{i=1}^{n}(-1)^{i+1}a_A(x_i)\varphi(x_1, \cdots,\hat{x_i},\cdots,x_{n+1})\nonumber\\
 &&-\sum_{i=1}^{n}(-1)^{i+1}\varphi(x_1, \cdots,\hat{x_i},\cdots,x_n,x_i\cdot_A x_{n+1})\nonumber\\
 &&+\sum_{1\leq i<j\leq n}(-1)^{i+j}\varphi([x_i,x_j]_A,x_1,\cdots,\hat{x_i},\cdots,\hat{x_j},\cdots,x_{n+1}).
\end{eqnarray}


\subsection*{Pre-symplectic algebroids}

Here we recall the notion of  pre-symplectic algebroids and the relation with symplectic Lie algebroids. See \cite{lsb} for more details.

 \begin{defi}\label{MTLA}
 A {\bf pre-symplectic algebroid} is a vector bundle $E\rightarrow M$ equipped with a nondegenerate skew-symmetric bilinear form $(\cdot,\cdot)_-$, a multiplication $\star:\Gamma(E)\times\Gamma(E)\longrightarrow\Gamma(E) $, and a bundle map $\rho:E\rightarrow TM$, such that  for all $e, e_1,e_2,e_3\in\Gamma(E),~~f\in C^\infty(M)$, the following conditions are satisfied:
\begin{itemize}
\item[$\rm(i)$] $(e_1,e_2,e_3)-(e_2,e_1,e_3)=\frac{1}{6}DT(e_1,e_2,e_3)$\label{associator};
\item[$\rm(ii)$]$\rho(e_1)(e_2,e_3)_-=(e_1{\star}e_2-\frac{1}{2}D(e_1,e_2)_-,e_3)_-+(e_2,[e_1,e_3]_E)_-,$\label{invariant}
   \end{itemize}
   where $(e_1,e_2,e_3)$ is the associator for the multiplication $\star$ given by \eqref{eq:associator}, $T:\Gamma(E)\times\Gamma(E)\times\Gamma(E) \longrightarrow \CWM$ is defined by
   \begin{equation}\label{T-equation}
   T(e_1,e_2,e_3)= (e_1\star e_2, e_3)_-+(e_1,e_2\star e_3)_- - (e_2\star e_1, e_3)_--(e_2,e_1\star e_3)_-,
   \end{equation}
 $D:C^\infty(M)\longrightarrow\Gamma(E)$ is defined by
  \begin{equation}\label{eq:defD}
 (Df,e)_-=\rho(e)(f),
 \end{equation}
 and the bracket $[\cdot,\cdot]_E:\wedge^2\Gamma(E)\longrightarrow \Gamma(E)$ is defined by
 \begin{equation}\label{eq:defbraE}
 [e_1,e_2]_E=e_1\star e_2-e_2\star e_1.
 \end{equation}
\end{defi}

We denote a pre-symplectic algebroid  by  $(E,\star,\rho,(\cdot,\cdot)_-)$.

\begin{thm}\label{symp 3}
Let $(E,\star,\rho,(\cdot,\cdot)_-)$ be a pre-symplectic algebroid. Then
 $(E,[\cdot,\cdot]_E,\rho,\omega=(\cdot,\cdot)_-)$ is a symplectic Lie algebroid.
\end{thm}

 Given a symplectic Lie algebroid $(E,[\cdot,\cdot]_E,\rho,\omega)$, define a multiplication $\star:\Gamma(E)\times\Gamma(E)\longrightarrow \Gamma(E)$  by
\begin{eqnarray}\label{LSCA bracket}
  e_1\star e_2={\omega^\sharp}^{-1}(\huaL_{e_1}\omega^\sharp(e_2)+\half \dM(\omega(e_1,e_2)))\quad \forall e_1,e_2\in \Gamma(E).
\end{eqnarray}

\begin{thm}\label{thm:Symplectic-LWX}
Let $(E,[\cdot,\cdot]_E,\rho,\omega)$ be a symplectic Lie algebroid. Then $(E,\star,\rho,(\cdot,\cdot)_-=\omega)$ is a pre-symplectic algebroid, and satisfies
\begin{eqnarray}\label{bracket}
[e_1,e_2]_E=e_1\star e_2-e_2\star e_1,\quad \forall e_1,e_2\in \Gamma(E),
\end{eqnarray}
where the multiplication $\star$ is given by $(\ref{LSCA bracket})$.
\end{thm}

\begin{ex} \label{ex:pseudosemidirectp}{\rm
  Let $(A,\cdot_A,a_A)$ be a left-symmetric algebroid and $(A^c\ltimes_{L^*} A^*,\omega)$ the corresponding symplectic Lie algebroid, where $\omega$ is given by \eqref{eq:defiomega}. Then the corresponding pre-symplectic algebroid structure is given by
  \begin{equation}\label{eq:standardLWXbracket}
    (x+\xi)\star(y+\eta)=x\cdot_A y+\huaL_x\eta-R_y\xi-\half \dM(x+\xi,y+\eta)_+,
  \end{equation}
where $\huaL$ is given by \eqref{Lie der1}, $R$ is given by \eqref{eq:Rformula}, and $(\cdot,\cdot)_+$ is the nondegenerate symmetric bilinear form on $A\oplus A^*$ given by
\begin{equation}\label{sym-form}
 (x+\xi,y+\eta)_+=\langle\xi,y\rangle+\langle\eta,x\rangle.
\end{equation}
}
\end{ex}

 \begin{defi}
 Let $(E,\star,\rho,(\cdot,\cdot)_-)$ be a pre-symplectic algebroid. A subbundle $F$ of $E$ is called {\bf isotropic}
  if it is isotropic under the skew-symmetric bilinear form $(\cdot,\cdot)_-$. It is called {\bf integrable}
  if $\Gamma(F)$ is closed under the operation $\star$. A {\bf Dirac structure} is a subbundle $F$ which is maximal isotropic and integrable.
 \end{defi}

The following proposition is obvious.
\begin{pro}\label{Dirac subbundles}
Let $F$ be a Dirac structure of a pre-symplectic algebroid $(E,\star,\rho,(\cdot,\cdot)_-)$. Then $(F,\star|_F,\rho|_F)$ is a left-symmetric algebroid.
\end{pro}

\section{Differential calculus on left-symmetric algebroids}
In this section, we develop the differential calculus on
left-symmetric algebroids, which is the fundamental tool in the
following study.
 Let $(A,\cdot_A, a_A)$ be a left-symmetric algebroid. 
For all $x\in\Gamma(A)$, define the {\bf Lie derivative} $\frkL_x: \Gamma(\wedge^{n}A\otimes A)\longrightarrow \Gamma(\wedge^{n}A\otimes A)$ by
\begin{eqnarray}\label{eq:LiederPre}
\frkL_x(y_1\wedge \cdots\wedge y_{n}\otimes y_{n+1})=\sum_{i=1}^{n}y_1\wedge \cdots \wedge
x\cdot_{A}y_i\wedge\cdots\wedge y_{n}\otimes y_{n+1} +y_1\wedge
\cdots\wedge y_{n}\otimes [x,y_{n+1}]_{A},
\end{eqnarray}
where $y_1, \cdots,y_{n+1}\in\Gamma(A)$. Define the {\bf right multiplication} $R_x: \Gamma(\wedge^{n}A\otimes A)\longrightarrow \Gamma(\wedge^{n}A\otimes A)$  by
\begin{eqnarray}
R_x(y_1\wedge \cdots\wedge y_{n}\otimes y_{n+1})=-(\sum_{i=1}^{n}y_1\wedge \cdots
y_i\cdot_{A}x\cdots\wedge y_{n}\otimes y_{n+1}) +y_1\wedge
\cdots\wedge y_{n}\otimes y_{n+1}\cdot_{A}x.
\end{eqnarray}

\begin{rmk}
  For all $x,y\in\Gamma(A)$, we have $\frkL_xy=[x,y]_A$. Thus, $\frkL_x$ is not a straightforward generalization of the left multiplication $L_x:\Gamma(A)\longrightarrow \Gamma(A)$ given by \eqref{eq:leftmul}.
  This is why we use different notations. However, $R_x:\Gamma(\wedge^{n}A\otimes A)\longrightarrow \Gamma(\wedge^{n}A\otimes A)$ is a straightforward generalization of the right multiplication $R_x:\Gamma(A)\longrightarrow \Gamma(A)$ given by \eqref{eq:leftmul}.
  Therefore, we use the same notations  which will not cause confusion.
\end{rmk}


For all $\xi\in\Gamma(A^*)$, the {\bf left contraction} and  {\bf right contraction}, which we denote by
$_{\xi}\lrcorner:\Gamma(\wedge^{n}A\otimes A)\longrightarrow \Gamma(\wedge^{n-1}A\otimes A)$ and $\llcorner_{\xi}:\Gamma(\wedge^{n}A\otimes A)\longrightarrow \Gamma(\wedge^{n}A)$ respectively,  are
defined by
\begin{eqnarray}
(_{\xi}\lrcorner\varphi)(\eta_1,\eta_2,\cdots,\eta_n)&=&\varphi(\xi,\eta_1,\eta_2,\cdots,\eta_n);\\
(\varphi\llcorner_{\xi})(\eta_1,\eta_2,\cdots,\eta_n)&=&\varphi(\eta_1,\eta_2,\cdots,\eta_n,\xi),
\end{eqnarray}
 where  $\varphi\in \Gamma(\wedge^{n}A\otimes A),\quad n\geq1$ and $\eta_1,\eta_2,\cdots,\eta_n\in \Gamma(A^*)$.

 The Lie derivative $\frkL_x$ has the following properties.
\begin{pro}
For all $x,y\in \Gamma(A),~~f\in\CWM,~~X\in \Gamma(\wedge^{n}A\otimes A),n\geq0$, we have
\begin{eqnarray}
\frkL_{[x,y]_A}&=&[\frkL_x,\frkL_y];\\
\label{pro 2 of lie der}\frkL_{x}fX&=&f\frkL_xX+a_A(x)(f)X;\\
\frkL_{fx}X&=&f\frkL_xX-X\llcorner_{\dM f}\otimes x.\label{pro 3 of lie der}
\end{eqnarray}
\end{pro}
\pf We only prove  (\ref{pro 3 of lie der}). Others can be proved similarly.
For all $x,y_1,y_2,\cdots,y_{n+1}\in \Gamma(A),~~f\in \CWM$, without loss of generality we can assume that $X=y_1\wedge\cdots\wedge y_n\otimes y_{n+1}$, then we have
\begin{eqnarray*}
&&\frkL_{fx}y_1\wedge\cdots\wedge y_n\otimes y_{n+1}\\
&=&\sum_{i=1}^{n}y_1\wedge \cdots \wedge
(fx)\cdot_{A}y_i\wedge\cdots\wedge y_{n}\otimes y_{n+1} +y_1 \wedge\cdots\wedge y_{n}\otimes [fx,y_{n+1}]_{A}\\
&=&f\sum_{i=1}^{n}y_1\wedge \cdots \wedge
x\cdot_{A}y_i\wedge\cdots\wedge y_{n}\otimes y_{n+1} +fy_1\wedge
 \cdots\wedge y_{n}\otimes [x,y_{n+1}]_{A}\\
&&-y_1\wedge\cdots\wedge y_{n}\otimes a_A(y_{n+1})(f)x\\
&=&f\frkL_{x}y_1\wedge\cdots\wedge y_n\otimes y_{n+1}-(y_1\wedge\cdots\wedge y_n\otimes y_{n+1})\llcorner_{\dM f}\otimes x.
\end{eqnarray*}
The proof is finished.\qed\vspace{3mm}

For all   $x\in \Gamma(A)$, the {\bf Lie derivative} $\frkL_x:\Gamma(\wedge^{n}A^*\otimes A^*)\longrightarrow \Gamma(\wedge^{n}A^*\otimes A^*)$ and the {\bf right multiplication} $R_x:\Gamma(\wedge^nA^*\otimes A^*)\longrightarrow \Gamma(\wedge^nA^*\otimes A^*)$ are defined respectively by\footnote{Here we use the same notations as before and this will not bring confusion since it depends on what it acts.}
\begin{eqnarray}
\label{Lie Der}\langle \frkL_x\varphi,X\rangle &=&a_A(x)\langle\varphi,X\rangle-\langle\varphi,\frkL_xX\rangle;\\
\label{eq:Rformula}\langle R_x\varphi,X\rangle&=&-\langle\varphi,R_xX\rangle.
\end{eqnarray}
where $\varphi\in \Gamma(\wedge^{n}A^*\otimes A^*),\quad n\geq1$  and $X\in \Gamma(\wedge^{n}A\otimes A)$.

\emptycomment{
Obviously, we have
\begin{pro}
For all $\varphi\in \Gamma(\wedge^nA^*\otimes A^*),\quad n\geq1$ and $x,y_1,y_2,\cdots,y_n\in \Gamma(A)$, we have
\begin{eqnarray}
&&(\frkL_x\varphi)(y_1,y_2,\cdots,y_{n+1})\nonumber\\
&&=a_A(x) \varphi(y_1,y_2,\cdots,y_{n+1})-\varphi(y_1,y_2,\cdots,[x,
y_{n+1}])-\sum_{i=1}^{n}\varphi(y_1,\cdots,x\cdot_A y_j,\cdots, y_{n+1});\\
&&(R_x\varphi),(y_1,y_2,\cdots,y_{n+1})\nonumber \\
&&=\sum_{i=1}^{n}\varphi(y_1, \cdots ,
y_i\cdot_{A}x,\cdots, y_{n}, y_{n+1}) -\varphi(y_1,
y_2,\cdots, y_{n}, y_{n+1}\cdot_{A}x).
\end{eqnarray}
\end{pro}
\comment{not needed}
}

These operators satisfy the following equalities which are repeatedly used below.
\begin{pro}
For all $\varphi\in \Gamma(\wedge^nA^*\otimes A^*),n\geq1,~~x,y\in \Gamma(A),\xi\in\Gamma(A^*)$, we have
\begin{eqnarray}
{\delta}(f\xi)&=&f{\delta}(\xi)+\dM f\otimes
\xi;\\
\frkL_{[x,y]_A}&=&[\frkL_x,\frkL_y];\\
_{(x\cdot_{A}y)}\lrcorner\varphi&=&\frkL_x(_{y}\lrcorner\varphi)-_{y}\lrcorner(\frkL_x\varphi);\label{formuar 3}\\
\frkL_x \varphi&=&\delta (_{x}\lrcorner\varphi)+(_{x}\lrcorner)\delta\varphi-R_x\varphi;\label{formuar 4}\\
\frkL_{x}f\varphi&=&f\frkL_x\varphi+a_A(x)(f)\varphi;\\
\frkL_{fx}\varphi&=&f\frkL_x\varphi+\varphi\llcorner_x\otimes\dM f;\\
R_{x}f\xi&=&fR_{x}\xi;\\
R_{fx}\xi&=&fR_{x}\xi-\langle x,\xi\rangle\dM f.
\end{eqnarray}
\end{pro}
\pf We only give the proof of  (\ref{formuar 3}) and (\ref{formuar 4}). Others can be proved similarly.
For all $\varphi\in \Gamma(\wedge^nA^*\otimes A^*),\quad n\geq1$ and $x,y,y_1,y_2,\cdots,y_n\in \Gamma(A)$, we have
\begin{eqnarray*}
&&\frkL_x(_{y}\lrcorner\varphi)(y_1,\cdots,y_n)\\
&=&a_A(x){_{y}\lrcorner\varphi}(y_1,\cdots,y_n)-{_{y}\lrcorner\varphi}(y_1,\cdots,[x,y_n]_A)-\sum_{i=1}^{n-1}{_{y}\lrcorner\varphi}(y_1,\cdots,x\cdot_A y_i,\cdots, y_{n})\\
&=&a_A(x)\varphi(y,y_1,\cdots,y_n)-\varphi(y,y_1,\cdots,[x,y_n]_A)-\sum_{i=1}^{n-1}\varphi(y,y_1,\cdots,x\cdot_A y_i,\cdots, y_{n})
\end{eqnarray*}
and
\begin{eqnarray*}
&&_{y}\lrcorner(\frkL_x\varphi)(y_1,\cdots,y_n)
=\frkL_x\varphi(y,y_1,\cdots,y_n)\\
&=&a_A(x)\varphi(y,y_1,\cdots,y_n)-\varphi(y,y_1,\cdots,[x,y_n]_A)-\sum_{i=1}^{n-1}\varphi(y,y_1,\cdots,x\cdot_A y_i,\cdots, y_{n})\\
&&-\varphi(x\cdot_A y,y_1,\cdots,y_n).
\end{eqnarray*}
Therefore, we have
\begin{eqnarray*}
\frkL_x(_{y}\lrcorner\varphi)(y_1,\cdots,y_n)-_{y}\lrcorner(\frkL_x\varphi)(y_1,\cdots,y_n)=\varphi(x\cdot_A y,y_1,\cdots,y_n),
\end{eqnarray*}
which implies that (\ref{formuar 3}) holds.

Next we prove that  (\ref{formuar 4}) holds. On  one hand, for all $\varphi\in \Gamma(\wedge^nA^*\otimes A^*),~n\geq1$ and $x,y_1,\cdots,y_{n+1}\in \Gamma(A)$, we have
\begin{eqnarray*}
 &&(_{x}\lrcorner)\delta\varphi(y_1,\cdots,y_{n+1})=\delta\varphi(x,y_1,\cdots,y_{n+1})\\
&=&a_A(x)\varphi(y_1, \cdots,y_{n+1})+\sum_{i=1}^{n}(-1)^{i}a_A(y_i)\varphi(x,y_1, \cdots,\hat{y_i},\cdots,y_{n+1})\\
&& -\varphi(y_1,\cdots,y_{n},x\cdot_A y_{n+1})-\sum_{i=1}^{n}(-1)^{i}\varphi(x,y_1,\cdots,\hat{y_i},\cdots,y_n,y_i\cdot_A y_{n+1})\\
&&+\sum_{i=1}^{n}(-1)^{i}\varphi([x,y_i]_A,y_1,\cdots,\hat{y_i},\cdots,y_{n+1}) \\
&& +\sum_{1\leq i<j\leq n}(-1)^{i+j}\varphi([y_i,y_j]_A,x,y_1,\cdots,\hat{y_i},\cdots,\hat{y_j},\cdots,y_{n+1})
\end{eqnarray*}
and
\begin{eqnarray*}
 \delta (_{x}\lrcorner\varphi)(y_1,\cdots,y_{n+1})&=&\sum_{i=1}^{n}(-1)^{i+1}a_A(y_i)\varphi(x,y_1,\cdots,\hat{y_i},\cdots,y_{n+1})\\
 &&-\sum_{i=1}^{n}(-1)^{i+1}\varphi(x,y_1,\cdots,\hat{y_i},\cdots,y_{n},y_i\cdot_A y_{n+1})\\
 &&+\sum_{1\leq i<j\leq n}(-1)^{i+j}\varphi(x,[y_i,y_j]_A,y_1,\cdots,\hat{y_i},\cdots,\hat{y_j},\cdots,y_{n+1}).
\end{eqnarray*}
On the other hand, by \eqref{Lie Der} and \eqref{eq:Rformula}, we have
\begin{eqnarray*}
&&(\frkL_x \varphi+R_x\varphi)(y_1,\cdots,y_{n+1})\\
&=&a_A(x)\varphi(y_1, \cdots,y_{n+1})-\varphi(y_1, \cdots,y_{n},x\cdot_A y_{n+1}) \\ &&+\sum_{i=1}^{n}(-1)^{i}\varphi([x,y_i]_A,y_1,\cdots,\hat{y_i},\cdots,y_{n+1}).
\end{eqnarray*}
Thus, (\ref{formuar 4}) holds.\qed




\section{Left-symmetric bialgebroids}
In this section, we introduce the concept of a left-symmetric bialgebroid and study its properties. We construct a left-symmetric algebroid using a symmetric tensor satisfying a condition, which can be viewed as a generalization of the S-equation introduced in \cite{Left-symmetric bialgebras} for left-symmetric bialgebras. In particular, we construct a left-symmetric bialgebroid  using a pseudo-Hessian manifold.

\begin{defi}
Let $(A,\cdot_A,a_A)$ and $(A^*,\cdot_{A^*},a_{A^*})$ be two
left-symmetric algebroids. Then $(A,A^*)$ is a {\bf left-symmetric
bialgebroid} if for all $x,y\in\Gamma(A),~~\xi,\eta\in\Gamma(A^*)$,
the following equalities hold:
\begin{eqnarray}
\delta[\xi,\eta]_{A^*}&=&\frkL_\xi{\delta}\eta-\frkL_\eta{\delta}\xi;\label{cond1}\\
\delta_*[x,y]_{A}&=&\frkL_x{\delta}_*y-\frkL_y{\delta}_*x\label{cond2},
\end{eqnarray}
where $\delta$ and $\delta_*$ are coboundary operators of left-symmetric algebroids  $A$ and $A^*$ respectively and $\frkL$ is the Lie derivative associated to a left-symmetric algebroid given by \eqref{eq:LiederPre}.
\end{defi}

\begin{rmk}
 By  \eqref{eq:LiederPre}, it is not hard to see that a left-symmetric bialgebroid reduces to a left-symmetric bialgebra when the base manifold is a point. Thus, a left-symmetric bialgebroid can be viewed as a geometric generalization of a left-symmetric bialgebra. See \cite{Left-symmetric bialgebras} for more details about left-symmetric bialgebras.
\end{rmk}


\begin{lem} Let $(A,A^*)$ be a left-symmetric bialgebroid.
For all $x\in\Gamma(A),~~ \xi\in\Gamma(A^*),~~f\in \CWM$, we have
\begin{eqnarray}
x\cdot_A \dM_*f&=&\dM_*(x\llcorner_{\dM
f})-({\delta}_*x)\llcorner_{\dM f},\label{pro of lsboids1}\\
 \xi\cdot_{A^*} \dM
f&=&\dM(x\llcorner_{\dM_*
f})-(\delta\xi)\llcorner_{\dM_* f}\label{pro of lsboids2}.
\end{eqnarray}
\end{lem}
\pf By \eqref{pro
2 of lie der}, $(\ref{pro 3 of lie der})$ and $(\ref{cond2})$, we have
\begin{eqnarray*}
{\delta}_*[x,fy]_{A}&=&\frkL_x{\delta}_*fy-\frkL_{fy}{\delta}_*x\\
&=&\frkL_x(f{\delta}_*y+\dM_*f\otimes
y)-(f\frkL_{y}{\delta}_*x-({\delta}_*x)\llcorner_{\dM
f}\otimes y)\\
&=&f\frkL_x{\delta}_*y+a_{A}(x)(f){\delta}_*y+\frkL_x(\dM_*f\otimes
y)-f\frkL_{y}{\delta}_*x+({\delta}_*x)\llcorner_{\dM
f}\otimes y\\
&=&f\delta_*[x,y]_A+a_{A}(x)(f){\delta}_*y+(x\cdot_A\dM_*f)\otimes
y+\dM_*f\otimes[x,y]_A+({\delta}_*x)\llcorner_{\dM
f}\otimes y.
\end{eqnarray*}
On the other hand, we have
\begin{eqnarray*}
{\delta}_*[x,fy]_{A}&=&{\delta}_*(f[x,y]_A+a_A(x)(f)y)\\
&=&f{\delta}_*[x,y]_A+\dM_*f\otimes
[x,y]_A+a_{A}(x)(f){\delta}_*y+\dM_*a_A(x)(f)\otimes y.
\end{eqnarray*}
Thus, we have
$$x\cdot_A
\dM_*f+({\delta}_*x)\llcorner_{\dM f}=\dM_*a_A(x)(f)=\dM_*(x\llcorner_{\dM
f}), $$
which implies that \eqref{pro of lsboids1} holds. \eqref{pro of lsboids2} can be proved similarly. \qed\vspace{3mm}

Recall that $R_x:A^*\longrightarrow A^*$ and
$R_\xi:A\longrightarrow A$ are defined by
$$\langle R_x \xi,y\rangle=-\langle\xi,y\cdot_{A}x\rangle,\quad \langle R_\xi x,\eta\rangle=-\langle x,\eta\cdot_{A^*}\xi\rangle.$$

\begin{cor} Let $(A,A^*)$ be a left-symmetric bialgebroid.
For all $x\in\Gamma(A),~~ \xi\in\Gamma(A^*),~~f\in \CWM$, we have
\begin{eqnarray}
x\cdot_A \dM_*f&=&- R_{\dM f}x,\label{pro of lsboids11}\\
\xi\cdot_{A^*} \dM
f&=&-R_{\dM_* f}\xi.\label{pro of lsboids12}
\end{eqnarray}
\end{cor}
\pf By (\ref{pro of lsboids1}), we have
\begin{eqnarray*}
\langle\dM_*(x\llcorner_{\dM
f})-({\delta}_*(x))\llcorner_{\dM
f},\xi\rangle&=&\langle\dM_*a_A(x)(f),\xi\rangle-\langle{\delta}_*(x),\xi\otimes\dM
f \rangle\\
&=&a_{A^*}(\xi)a_A(x)(f)-a_{A^*}(\xi)a_A(x)(f)+\langle
x,\xi\cdot_{A^*}\dM f\rangle\\
&=&\langle-R_{\dM f}x,\xi\rangle,
\end{eqnarray*}
which implies that (\ref{pro of lsboids11}) holds. (\ref{pro of lsboids12}) can be proved similarly.\qed

\begin{cor}\label{pro2}
Let $(A,A^*)$ be a left-symmetric bialgebroid.
For all $x\in\Gamma(A),~~\xi\in\Gamma(A^*)$, we have
\begin{equation}\label{eq:formula}
[a_A(x),a_{A^*}(\xi)]=a_{A^*}({L}^*_x\xi)-a_{A}({L}^*_\xi x).
\end{equation}
\end{cor}
\pf For all $f\in\CWM$, by \eqref{pro of lsboids11}, we have
\begin{eqnarray*}
\langle a_{A^*}({L}^*_x\xi)-a_{A}({L}^*_\xi x),df\rangle&=&\langle \dM_* f,{L}^*_x\xi\rangle-\langle\dM f,{L}^*_\xi x\rangle\\
&=&a_A(x)\langle\dM_*f,\xi\rangle-\langle x\cdot_A \dM_* f,\xi\rangle-a_{A^*}(\xi)\langle\dM f,x\rangle+\langle\xi\cdot_{A^*}\dM f,x\rangle\\
&=&a_A(x)\langle\dM_*f,\xi\rangle-a_{A^*}(\xi)\langle\dM f,x\rangle+\langle R_{\dM f}x,\xi\rangle+\langle\xi\cdot_{A^*}\dM f,x\rangle\\
&=&\langle[a_A(x),a_{A^*}(\xi)],d f\rangle,
\end{eqnarray*}
which implies that \eqref{eq:formula} holds. \qed\vspace{3mm}

Let $(A,\cdot_A,a_A)$ be a left-symmetric algebroid. Define
$$\Sym^2(A)=\{H\in A\otimes A| H(\xi,\eta)=H(\eta,\xi),\quad \forall\xi,\eta\in \Gamma(A^*)\}.$$
For any $H\in \Sym^2(A)$, the bundle map  $H^\sharp:A^*\longrightarrow A$ is given by $H^\sharp(\xi)(\eta)=H(\xi,\eta)$.
 We introduce   $\llbracket H,H\rrbracket\in\wedge^2A \otimes A $ as follows:
\begin{eqnarray}\label{brac2}
\llbracket H,H\rrbracket(\xi_1,\xi_2,\xi_3)&=&a_A(H^\sharp(\xi_1))\langle H^\sharp(\xi_2),\xi_3\rangle-a_A(H^\sharp(\xi_2))\langle H^\sharp(\xi_1),\xi_3\rangle+\langle \xi_1,H^\sharp(\xi_2)\cdot_A H^\sharp(\xi_3)\rangle \nonumber \\
&&-\langle\xi_2,H^\sharp(\xi_1)\cdot_A H^\sharp(\xi_3)\rangle-\langle \xi_3,[H^\sharp(\xi_1),
H^\sharp(\xi_2)]_A\rangle,\quad \forall \xi_1,\xi_2,\xi_3\in\Gamma(A^*).
\end{eqnarray}

 Suppose that $H^\sharp:A^*\longrightarrow A$ is   nondegenerate. Then   $(H^\sharp)^{-1}:A\longrightarrow A^*$ is also a symmetric bundle map, which gives rise to an element, denoted by $H^{-1}$,  in $\Sym^2(A^*)$.
\begin{pro}\label{pro:equivelent}
Let $(A,\cdot_A,a_A)$ be a left-symmetric algebroid and $H\in \Sym^2(A)$. If $H$ is nondegenerate, then the following two statements are equivalent:
\begin{itemize}
\item[$\rm(1)$]$\llbracket H,H\rrbracket=0$;
\item[$\rm(2)$]$\delta (H^{-1})=0.$
\end{itemize}
\end{pro}
\pf By direct calculation, we have the following formula
$$\delta (H^{-1})(H^\sharp(\xi_1),H^\sharp(\xi_2),H^\sharp(\xi_3))=\llbracket H,H\rrbracket(\xi_1,\xi_2,\xi_3),\quad \forall \xi_1,\xi_2,\xi_3\in\Gamma(A^*).$$
Thus, the conclusion follows immediately. \qed\vspace{3mm}

Let $(A,\cdot_A,a_A)$ be a left-symmetric algebroid, and $H\in \Sym^2(A)$. Define
\begin{equation}\label{eq:multiplication-H}
\xi\cdot_H \eta=\huaL_{H^\sharp(\xi)}\eta-R_{H^\sharp(\eta)}\xi-\dM (H(\xi,\eta)), \quad\forall \xi,\eta\in\Gamma(A^*).
\end{equation}

\begin{pro}With the above notations,
for all $\xi,\eta\in\Gamma(A^*)$, we have
\begin{equation}\label{homo}
H^\sharp(\xi\cdot_H \eta)-H^\sharp(\xi)\cdot_A H^\sharp(\eta)=\llbracket H,H\rrbracket(\xi,\cdot,\eta).
\end{equation}
\end{pro}
\pf First, for all $\xi,\eta,\zeta\in\Gamma(A^*)$, we have
\begin{eqnarray*}
\langle H^\sharp(\xi\cdot_H \eta),\zeta\rangle&=&\langle \xi\cdot_H \eta, H^\sharp(\zeta)\rangle=\langle\huaL_{H^\sharp(\xi)}\eta-R_{H^\sharp(\eta)}\xi-\dM (H(\xi,\eta)), H^\sharp(\zeta)\rangle\\
&=&a_A(H^\sharp(\xi))\langle H^\sharp(\zeta),\eta\rangle-a_A(H^\sharp(\zeta))\langle H^\sharp(\xi),\eta\rangle+\langle \xi,H^\sharp(\zeta)\cdot_A H^\sharp(\eta)\rangle \\
&&-\langle \eta,[H^\sharp(\xi),H^\sharp(\zeta)]\rangle.
\end{eqnarray*}
Thus, by (\ref{brac2}), we have
\begin{eqnarray*}
\langle H^\sharp(\xi\cdot_H \eta),\zeta\rangle-\langle H^\sharp(\xi)\cdot_A H^\sharp(\eta),\zeta\rangle=\llbracket H,H\rrbracket(\xi,\zeta,\eta),
\end{eqnarray*}
which finishes the proof.\qed\vspace{3mm}

By direct calculation, we have
\begin{cor}
For all $\xi,\eta\in\Gamma(A^*)$, we have
\begin{equation}\label{homo1}
H^\sharp({[\xi,\eta]}_H)-[H^\sharp(\xi), H^\sharp(\eta)]_A=\llbracket H,H\rrbracket(\xi,\eta,\cdot),
\end{equation}
where ${[\cdot,\cdot]}_H$ is the commutator
 bracket of $\cdot_H$.
\end{cor}

\begin{thm}\label{thm:LSBi-H}
With the above notations, if $\llbracket H,H\rrbracket=0$, then $(A^*,\cdot_H,a_{A^*}=a_A\circ H^\sharp)$ is a left-symmetric algebroid, and $H^\sharp$ is a left-symmetric algebroid homomorphism from $(A^*,\cdot_H,a_{A^*})$ to $(A,\cdot_A,a_A)$. Furthermore, $(A,A^*)$ is a left-symmetric bialgebroid.
\end{thm}
\pf By direct calculation, we can show that $(\Gamma(A^*),\cdot_H)$
is a left-symmetric algebra. Moreover, we have
\begin{eqnarray*}
\xi\cdot_H f\eta&=&(f\huaL_{H^\sharp\xi}\eta+a_A\circ H^\sharp(\xi)(f)\eta)-(fR_{H^\sharp\eta}\xi-\langle H^\sharp(\xi),\eta\rangle\dM f)-(f\dM H(\xi,\eta)+H(\xi,\eta)\dM f)\\
&=&f(\xi\cdot_H \eta)+a_A\circ H^\sharp(\xi)(f)\eta;\\
(f\xi)\cdot_H \eta&=&(f\huaL_{H^\sharp\xi}\eta+ \langle H^\sharp(\xi),\eta\rangle\dM f)-fR_{H^\sharp\eta}\xi-(f\dM H(\xi,\eta)+H(\xi,\eta)\dM f)\\
&=&f(\xi\cdot_H \eta).
\end{eqnarray*}
Thus, $(A^*,\cdot_H,a_A\circ H)$ is a left-symmetric algebroid.
 By $(\ref{homo})$, $H^\sharp$ is a left-symmetric algebroid homomorphism.

To obtain that $(A,A^*)$ is a left-symmetric bialgebroid, we need to prove that $(\ref{cond1})$ and $(\ref{cond2})$ hold. By tedious calculation,   $(\ref{cond1})$ can be obtained directly and    $(\ref{cond2})$ is equivalent to the following equation:
\begin{eqnarray}\label{homo2}
a_A[H^\sharp(\xi),x]=a_A\big(L^*_\xi x-H^\sharp(L^*_x\xi)\big).
\end{eqnarray}
 By direct calculation,
we have
\begin{eqnarray*}
\langle L^*_\xi x-H^\sharp(L^*_x\xi),\eta\rangle&=&a_{A^*}(\xi)\langle x,\eta\rangle-\langle x,\xi\cdot_H \eta\rangle-a_A(x)\langle \xi,H^\sharp(\eta)\rangle+\langle \xi,x\cdot_AH^\sharp(\eta)\rangle\\
&=&a_{A^*}(\xi)\langle x,\eta\rangle-a_A(H^\sharp(\xi))\langle x,\eta\rangle+a_A(x)\langle H^\sharp(\xi),\eta\rangle+\langle \eta,[H^\sharp(\xi),x]\rangle\\
&&-\langle \xi,x\cdot_A H^\sharp(\eta)\rangle-a_A(x)\langle \xi,H^\sharp(\eta)\rangle+\langle \xi,x\cdot_AH^\sharp(\eta)\rangle\\
&=&\langle[H^\sharp(\xi),x],\eta\rangle,
\end{eqnarray*}
which implies that $(\ref{homo2})$ holds.\qed

\begin{rmk}
When $M$ is a point, $\llbracket H,H\rrbracket=0$ is exactly the S-equation introduced in \cite{Left-symmetric bialgebras}.
Thus, the equation $\llbracket H,H\rrbracket=0$ can be viewed as a geometric
 generalization of the S-equation.
\end{rmk}

At the end of this section, we give some applications in Hessian geometry and construct a left-symmetric bialgebroid from a pseudo-Hessian manifold.

Recall that a {\bf pseudo-Hessian metric} $g$ is a
pseudo-Riemannian metric $g$ on a flat manifold $(M,\nabla)$ such
that $g$ can be locally expressed by
$g_{ij}=\frac{\partial^2\varphi}{\partial x^i\partial x^j},$ where
$\varphi\in\CWM$ and $\{x^1,\cdots,x^n\}$ is an affine coordinate
system with respect to $\nabla$, i.e.
$$\nabla_{\frac{\partial}{\partial x^i}}{\frac{\partial}{\partial
x^i}}=0,~i=1,\ldots,n.$$ Then the pair $(\nabla,g)$ is called a
pseudo-Hessian structure on $M$. A manifold $M$ with a
pseudo-Hessian structure $(\nabla,g)$ is called a  {\bf pseudo-Hessian manifold.}
 See \cite{Geometry of Hessian structures} for more
details about pseudo-Hessian manifolds.
\begin{pro}\label{Hessian}{\rm\cite{Geometry of Hessian structures}}
Let $(M,\nabla)$ be a flat manifold and $g$ a pseudo-Riemannian metric on $M$. Then the following conditions are equivalent:
\begin{itemize}
\item[$\rm(1)$]$\g$ is a pseudo-Hessian metric;
\item[$\rm(2)$] for all $x,y,z\in\Gamma(TM),$ there holds $\nabla_xg(y,z)=\nabla_yg(x,z),$
where $\nabla_xg(y,z)$ is given by
\begin{eqnarray*}
\nabla_xg(y,z)=xg(y,z)-g(\nabla_x y,z)-g(y,\nabla_xz).
\end{eqnarray*}
\end{itemize}
\end{pro}

\begin{pro}\label{pro:hessian-cocycle}
Let $(M,\nabla)$ be a flat manifold and $g$ a pseudo-Riemannian metric on $M$. Then $(M,\nabla,g)$ is a pseudo-Hessian manifold if and only if $\delta g(x,y,z)=0$, where $\delta$ is the coboundary operator given by \eqref{LSCA cohomology} associated to the left-symmetric algebroid $T_\nabla M $ given in Example \ref{ex:main}.
\end{pro}
\pf By Proposition \ref{Hessian} and the following formula  $$\delta g(x,y,z)=\nabla_xg(y,z)-\nabla_yg(x,z),$$
we can obtain the conclusion   immediately. \qed\vspace{3mm}

By  Proposition \ref{pro:hessian-cocycle}, Proposition \ref{pro:equivelent} and Theorem \ref{thm:LSBi-H}, we can construct a left-symmetric bialgebroid from a pseudo-Hessian manifold.

\begin{pro}\label{cor:hessianLA}
Let $(M,\nabla,g)$ be a pseudo-Hessian manifold.  Define $H\in \Sym^2(TM)$ by
$$(H^\sharp)^{-1}(x)(y)=g(x,y).$$ Then  $(T^*M,\cdot_H, H^\sharp)$ is a left-symmetric algebroid, which we denote by $T^*_HM$, where $\cdot_H$ is given by \eqref{eq:multiplication-H}. Furthermore,   $H^\sharp$ is a left-symmetric algebroid homomorphism from $T^*_HM $ to $T_\nabla M$ and $(T_\nabla M,T^*_HM)$ is a left-symmetric bialgebroid.
\end{pro}

\begin{rmk}
  The above result is parallel to that $(TM,T^*_\pi M)$ is a Lie bialgebroid for any Poisson manifold $(M,\pi)$. See \cite{KS} for more details.
\end{rmk}

\section{Equivalence between Manin triples for left-symmetric algebroids and left-symmetric bialgebroids}
 In this section, we introduce the
notion of a Manin triple for left-symmetric algebroids as a
geometric generalization of a Manin triple for left-symmetric
algebras. We would like to point out that unlike the latter, the
double structure of the former is a pre-symplectic algebroid rather
than a left-symmetric algebroid. We show that Manin triples for
left-symmetric algebroids are equivalent to left-symmetric
bialgebroids. At the end of this section, we establish a relation
between Maurer-Cartan type equations and Dirac structures of the
pre-symplectic algebroid which is the corresponding double
structure for a left-symmetric bialgebroid.

\begin{defi}
A  {\bf Manin triple for left-symmetric algebroids}
is a triple $(E;L_1,L_2)$, where $E$ is a pre-symplectic algebroid, $L_1$ and $L_2$ are transversal Dirac structures.
\end{defi}

 \begin{ex}{\rm
Let $(A,\cdot_A,a_A)$ be a left-symmetric algebroid and $(A\oplus A^*,\star,\rho,(\cdot,\cdot)_-)$ the corresponding pre-symplectic algebroid given in Example \ref{ex:pseudosemidirectp}. Then $(A\oplus A^*;A,A^*)$ is a Manin triple for left-symmetric algebroids.
}
 \end{ex}

 More generally, we have

\begin{thm}\label{thm:equivalence2}
  There is a one-to-one correspondence between  Manin triples for left-symmetric algebroids and left-symmetric bialgebroids.
\end{thm}
\pf Follows form the following Proposition  \ref{pro:LSbi-Manin} and Proposition \ref{pro:Manin-LSbi}.\qed\vspace{3mm}

Suppose that both $(A,\cdot_A,a_A)$ and $(A^*,\cdot_{A^*},a_{A^*})$ are left-symmetric algebroids. Let $E=A\oplus A^*$. We introduce a multiplication $\star:\Gamma(E)\times\Gamma(E)\longrightarrow \Gamma(E)$  by
\begin{equation}\label{operation}
e_1\star e_2=(x_1\cdot_A x_2+\huaL_{\xi_1}x_2-R_{\xi_2}x_1-\frac{1}{2}\dM_*(e_1,e_2)_+)+(\xi_1\cdot_{A^*} \xi_2+\huaL_{x_1}\xi_2-R_{x_2}\xi_1-\frac{1}{2}\dM(e_1,e_2)_+),
\end{equation}
where $e_1=x_1+\xi_1,e_2=x_2+\xi_2$, $\huaL$ is given by (\ref{Lie der1}), $R$ is given by \eqref{eq:Rformula} and $(\cdot,\cdot)_+$ is given by \eqref{sym-form}.
Let $\rho:E\rightarrow TM$ be the bundle map defined by $\rho=a_A+a_{A^*}$. That is,
\begin{equation}\label{anchor}
\rho(x+\xi)=a_A(x)+a_{A^*}(\xi),~~~\forall x\in\Gamma(A),~~\xi\in\Gamma(A^*).
\end{equation}
It is obvious  that in this case the operator $D$ (see \eqref{eq:defD}) is given by
$$D=\dM-\dM_*,$$
where $\dM:C^\infty(M)\rightarrow \Gamma({A^*})$ and $\dM_*:C^\infty(M)\rightarrow \Gamma(A)$ are the usual differential operators associated to the sub-adjacent Lie algebroids $A^c$ and ${A^*}^c$ respectively.

\begin{pro}\label{pro:LSbi-Manin}
With the above notations, let $(A,A^*)$ be a left-symmetric bialgebroid. Then $(A\oplus A^*,\star,\rho,(\cdot,\cdot)_-)$ is a pre-symplectic algebroid, where the multiplication $\star$ is given by $(\ref{operation})$, $\rho=a_A+a_{A^*},$ and $(\cdot,\cdot)_-$ is given by \eqref{eq:defiomega}.
\end{pro}

Conversely, we have
\begin{pro}\label{pro:Manin-LSbi}
Let $(E,\star,\rho,(\cdot,\cdot)_-)$ be a pre-symplectic algebroid. Suppose that $L_1$ and $L_2$ are Dirac subbundles transversal to each other. Then $(L_1,L_2)$ is a left-symmetric bialgebroid, where $L_2$ is considered as the dual bundle of $L_1$ under the nondegenerate bilinear form $(\cdot,\cdot)_-$.
\end{pro}

We prove some lemmas first.
\begin{lem}\label{associator 1}
Let $(A,\cdot_A,a_A)$ and $(A^*,\cdot_{A^*},a_{A^*})$ be two left-symmetric algebroids. For all $e_i=x_i+\xi_i\in\Gamma(E), ~i=1,2,3,4$, we have
\begin{eqnarray*}
{\big(}(x_1,x_2,\xi_3)-(x_2,x_1,\xi_3),x_4+\xi_4{\big)}_-=\frac{1}{6}\big( DT(x_1,x_2,\xi_3),x_4+\xi_4\big)_--I_1-I_2+I_3,
\end{eqnarray*}
where $I_1,I_2,I_3$ is defined by
\begin{eqnarray*}
I_1&=&\langle\delta_*[x_1,x_2]_A-\frkL_{x_1}{\delta}_*{x_2}+\frkL_{x_2}{\delta}_*{x_1},\xi_4\otimes\xi_3\rangle;\\
I_2&=&\big([a_A(x_1),a_{A^*}(\xi_4)]-a_{A^*}({L}^*_{x_1}\xi_4)+a_{A}({L}^*_{\xi_1}x_4)\big)\langle x_2,\xi_3\rangle;\\
I_3&=&\big([a_A(x_2),a_{A^*}(\xi_4)]-a_{A^*}({L}^*_{x_2}\xi_4)+a_{A}({L}^*_{\xi_4}x_2)\big)\langle x_1,\xi_3\rangle.
\end{eqnarray*}
\end{lem}
\pf
First, we have
\begin{eqnarray*}
&&(x_1,x_2,\xi_3)-(x_2,x_1,\xi_3)\\
&=&\huaL_{x_1}\huaL_{x_2}\xi_3-\huaL_{x_2}L^*_{x_1}\xi_3-\huaL_{[x_1,x_2]_{A}}\xi_3+\big(-x_1\cdot_{A}R_{\xi_3}x_2+x_2\cdot_{A}R_{\xi_3}x_1\\
&&-R_{\huaL_{x_2}\xi_3}x_1+R_{L^*_{x_1}\xi_3}x_2+R_{\xi_3}[x_1,x_2]_A\big)-\frac{1}{2}x_1\cdot_A\dM_*\langle x_2,\xi_3\rangle+\frac{1}{2}x_2\cdot_A\dM_*\langle
x_1,\xi_3\rangle\\
&&-\frac{1}{2}\dM_*\langle x_1,\huaL_{x_2}\xi_3\rangle+\frac{1}{2}\dM_*\langle x_2,\huaL_{x_1}\xi_3\rangle+\frac{1}{2}R_{\dM\langle x_2,\xi_3\rangle}x_1-\frac{1}{2}R_{\dM\langle x_1,\xi_3\rangle}x_2+\frac{1}{2}\dM_*\langle[x_1,x_2],\xi_3\rangle\\
&&+\frac{1}{4}\dM_*\langle x_1,\dM\langle x_2,\xi_3\rangle\rangle-\frac{1}{4}\dM_*\langle x_2,\dM\langle x_1,\xi_3\rangle\rangle-\frac{1}{2}\huaL_{x_1}\dM\langle x_2,\xi_3\rangle+\frac{1}{2}\huaL_{x_2}\dM\langle x_1,\xi_3\rangle\\
&&-\frac{1}{2}\dM\langle x_1,\huaL_{x_2}\xi_3\rangle+\frac{1}{2}\dM\langle x_2,\huaL_{x_1}\xi_3\rangle-\frac{1}{4}\dM\langle x_2,\dM\langle x_1,\xi_3\rangle\rangle+\frac{1}{4}\dM\langle x_1,\dM\langle x_2,\xi_3\rangle\rangle+\frac{1}{2}\dM\langle [x_1,x_2]_{A},\xi_3\rangle.
\end{eqnarray*}
By direct calculation, we have
\begin{eqnarray*}
&&\langle\delta_*[x_1,x_2]_A-\frkL_{x_1}{\delta}_*{x_2}+\frkL_{x_2}{\delta}_*{x_1},\xi_4\otimes\xi_3\rangle\\
&=&\langle-x_1\cdot_{A}R_{\xi_3}x_2+x_2\cdot_{A}R_{\xi_3}x_1-R_{\huaL_{x_2}\xi_3}x_1+R_{\huaL_{x_1}\xi_3}x_2+R_{\xi_3}[x_1,x_2]_A,\xi_4\rangle\\
&&+a_{A^*}(\xi_4)\langle[x_1,x_2]_A,\xi_3\rangle+([a_A(x_1),a_{A^*}(\xi_4)]+a_{A^*}({L}^*_{x_1}\xi_4))\langle x_2,\xi_3\rangle\\
&&-([a_A(x_2),a_{A^*}(\xi_4)]+a_{A^*}({L}^*_{x_2}\xi_4))\langle x_1,\xi_3\rangle.
\end{eqnarray*}
Therefore, we  obtain
\begin{eqnarray*}
&&{\big(}(x_1,x_2,\xi_3)-(x_2,x_1,\xi_3),x_4+\xi_4{\big)}_-\\
&=&-\langle\delta_*[x_1,x_2]_A-\frkL_{x_1}{\delta}_*{x_2}+\frkL_{x_2}{\delta}_*{x_1},\xi_4\otimes\xi_3\rangle\\
&&-\big([a_A(x_1),a_{A^*}(\xi_4)]-a_{A^*}({L}^*_{x_1}\xi_4)+a_{A}({L}^*_{\xi_1}x_4)\big)\langle x_2,\xi_3\rangle\\
&&+\big([a_A(x_2),a_{A^*}(\xi_4)]-a_{A^*}({L}^*_{x_2}\xi_4)+a_{A}({L}^*_{\xi_4}x_2)\big)\langle x_1,\xi_3\rangle\\
&&+\frac{1}{4}\big((\dM-\dM_*)\big(a_A(x_1)\langle x_2,\xi_3\rangle-a_A(x_2)\langle x_1,\xi_3\rangle-2\langle[x_1,x_2]_A,\xi_3\rangle\big),x_4+\xi_4\big)_-.
\end{eqnarray*}
It is easy to see that
\begin{eqnarray*}
&&\big(\frac{1}{6}\langle DT(x_1,x_2,\xi_3),x_4+\xi_4\big)_-\\
&=&\frac{1}{4}\big((\dM-\dM_*)\big(a_A(x_1)\langle x_2,\xi_3\rangle-a_A(x_2)\langle x_1,\xi_3\rangle-2\langle[x_1,x_2]_A,\xi_3\rangle\big),x_4+\xi_4\big)_-.
\end{eqnarray*}
The proof is finished.\qed

\begin{lem}\label{associator 2}
With the above notations, we have
\begin{eqnarray*}
\big((x_1,\xi_2,x_3)-(\xi_2,x_1,x_3),x_4+\xi_4\big)_-=\frac{1}{6}\big( DT(x_1,\xi_2,x_3),x_4+\xi_4\big)+J_1-J_2-J_3,
\end{eqnarray*}
where $J_1,J_2,J_3$ is defined by
\begin{eqnarray*}
J_1&=&\langle\delta_*[\xi_2,\xi_4]_A-\frkL_{\xi_2}{\delta}_*{\xi_4}+\frkL_{\xi_4}{\delta}_*{\xi_2},x_1\otimes x_3\rangle;\\
J_2&=&\big([a_{A^*}(\xi_2),a_{A}(x_1)]-a_{A}({L}^*_{\xi_2}x_1)+a_{A^*}({L}^*_{x_1}\xi_2)\big)\langle \xi_4,x_3\rangle;\\
J_3&=&\frac{1}{2}\big([a_{A^*}(\xi_4),a_{A}(x_1)]-a_{A}({L}^*_{\xi_4}x_1)+a_{A^*}({L}^*_{x_1}\xi_4)\big)\langle \xi_2,x_3\rangle.
\end{eqnarray*}
\end{lem}
\pf
First we have
\begin{eqnarray*}
&&(x_1,\xi_2,x_3)-(\xi_2,x_1,x_3)\\
&=&\big(-\huaL_{x_1}R_{x_3}\xi_2+R_{x_3}\huaL_{x_1}\xi_2+R_{x_1\cdot_A x_3}\xi_3+R_{x_3}R_{x_1}\xi_2\big)+x_1\cdot_{A}\huaL_{\xi_2}x_3+R_{R_{x_3}\xi_2}x_1\\
&&-\huaL_{\huaL_x{\xi_2}}x_3+(R_{\xi_2}x_1)\cdot_{A}x_3-\huaL_{\xi_2}(x_1\cdot_{A}x_3)+\huaL_{\xi_2}x_1\cdot_{A}x_3-\huaL_{R_{x_1}\xi_2}x_3\\
&&-\frac{1}{2}x_1\cdot_A\dM_*\langle\xi_2,x_3\rangle+\frac{1}{2}R_{\dM\langle\xi_2,x_3\rangle}x_1+\frac{1}{2}\dM_*\langle \huaL_{x_1}\xi_2,x_3\rangle+\frac{1}{2}\dM_*\langle R_{x_1}\xi_2,x_3\rangle\\
&&-\frac{1}{2}\huaL_{\dM\langle\xi_2,x_1\rangle}x_3+\frac{1}{4}\dM_*\langle x_1,\dM\langle\xi_2,x_3\rangle\rangle-\frac{1}{2}\huaL_{x_1}\dM\langle\xi_2,x_3\rangle\\
&&+\frac{1}{2}\dM\langle \huaL_{x_1}\xi_2,x_3\rangle+\frac{1}{2}\dM\langle R_{x_1}\xi_2,x_3\rangle+\frac{1}{4}\dM\langle x_1,\dM\langle\xi_2,x_3\rangle\rangle.
\end{eqnarray*}
Similarly, by direct calculation, we have
\begin{eqnarray*}
&&\langle\delta[\xi_2,\xi_4]_A-\frkL_{\xi_1}{\delta}{\xi_4}+\frkL_{\xi_4}{\delta}{\xi_2},x_1\otimes x_3\rangle\\
&=&-\langle x_1\cdot_{A}\huaL_{\xi_2}x_3+R_{R_{x_3}\xi_2}x_1-\huaL_{\huaL_x{\xi_2}}x_3+(R_{\xi_2}x_1)\cdot_{A}x_3-\huaL_{\xi_2}(x_1\cdot_{A}x_3),\xi_4\rangle\\
&&+\big([a_{A^*}(\xi_2),a_{A}(x_1)]-a_{A}({L}^*_{\xi_2}x_1)+a_{A^*}({L}^*_{x_1}\xi_2)\big)\langle \xi_4,x_3\rangle+a_{A^*}(\xi_4)\langle\xi_2,x_1\cdot_A x_3\rangle.
\end{eqnarray*}
Therefore, we obtain
\begin{eqnarray*}
&&\big((x_1,\xi_2,x_3)-(\xi_2,x_1,x_3),x_4+\xi_4\big)_-\\
&=&+\langle\delta[\xi_2,\xi_4]_A-\frkL_{\xi_1}{\delta}{\xi_4}+\frkL_{\xi_4}{\delta}{\xi_2},x_1\otimes x_3\rangle\\
&&-\big([a_{A^*}(\xi_2),a_{A}(x_1)]-a_{A}({L}^*_{\xi_2}x_1)+a_{A^*}({L}^*_{x_1}\xi_2)\big)\langle \xi_4,x_3\rangle\\
&&-\frac{1}{2}\big([a_{A^*}(\xi_4),a_{A}(x_1)]-a_{A}({L}^*_{\xi_4}x_1)+a_{A^*}({L}^*_{x_1}\xi_4)\big)\langle \xi_2,x_3\rangle\\
&&+\frac{1}{4}\big((\dM-\dM_*)\big(a_A(x_1)\langle \xi_2,x_3\rangle-2\langle x_1\cdot_A x_3,\xi_2\rangle,x_4+\xi_4\big)_-.
\end{eqnarray*}
It is easy to see that
$$\frac{1}{6}\big( DT(x_1,\xi_2,x_3),x_4+\xi_4\big)_-=\frac{1}{4}\big((\dM-\dM_*)\big(a_A(x_1)\langle \xi_2,x_3\rangle-2\langle x_1\cdot_A x_3,\xi_2\rangle\big),x_4+\xi_4\big)_-.$$
The proof is finished.\qed

\begin{lem}\label{associator 3}
With the above notations, we have
\begin{eqnarray*}
\big(\xi_1,x_2,x_3)-(x_2,\xi_1,x_3),x_4+\xi_4\big)_-=\frac{1}{6}\big( DT(\xi_1,x_2,x_3)_-,x_4+\xi_4\big)-J_1+J_2+J_3,
\end{eqnarray*}
where $J_1,J_2,J_3$ is defined by
\begin{eqnarray*}
J_1&=&\langle\delta_*[\xi_1,\xi_4]_A-\frkL_{\xi_1}{\delta}_*{\xi_4}+\frkL_{\xi_4}{\delta}_*{\xi_1},x_1\otimes x_3\rangle;\\
J_2&=&\big([a_{A^*}(\xi_1),a_{A}(x_2)]-a_{A}({L}^*_{\xi_1}x_2)+a_{A^*}({L}^*_{x_2}\xi_1)\big)\langle \xi_4,x_3\rangle;\\
J_3&=&\frac{1}{2}\big([a_{A^*}(\xi_4),a_{A}(x_2)]-a_{A}({L}^*_{\xi_4}x_2)+a_{A^*}({L}^*_{x_2}\xi_4)\big)\langle \xi_1,x_3\rangle.
\end{eqnarray*}
\end{lem}
\pf By Lemma $\ref{associator 2}$ and $T(e_1,e_2,e_3)=-T(e_2,e_1,e_3)$. The lemma follows immediately.\qed\vspace{3mm}

{\bf Proof of Proposition  \ref{pro:LSbi-Manin} } To prove
Proposition  \ref{pro:LSbi-Manin}, it is
sufficient to verify that conditions (i) and (ii) in Definition
$\ref{MTLA}$ hold. First, condition (i) in Definition \ref{MTLA}
follows directly from Lemma $\ref{associator 1}-\ref{associator
3}$ and properties of left-symmetric bialgebroids. Below, we show
that condition (ii) in Definition \ref{MTLA} holds. On one hand,
we have
\begin{eqnarray*}
\rho(e_1)(e_2,e_3)_-=(a_A(x_1)+a_{A^*}(\xi_1))(\langle\xi_2,x_3\rangle-\langle x_2,\xi_3\rangle),\quad \forall e_i=x_i+\xi_i\in\Gamma(E), ~i=1,2,3.
\end{eqnarray*}
On the other hand, we have
\begin{eqnarray*}
(e_1{\star}e_2-\frac{1}{2}D(e_1,e_2)_-,e_3)_-
&=&\langle \xi_1\cdot_{A^*}\xi_2,x_3\rangle-\langle\xi_2,[x_1,x_3]_A\rangle+\langle\xi_1,x_3\cdot_{A}x_2\rangle-\langle\xi_3,x_1\cdot_{A}x_2\rangle\\
&&+\langle x_2,[\xi_1,\xi_3]_A\rangle-\langle \xi_3\cdot_{A^*}\xi_2,x_1\rangle+a_A(x_1)\langle \xi_2,x_3\rangle\\
&&-a_{A^*}(\xi_1)\langle x_2,\xi_3\rangle-a_A(x_3)\langle \xi_1,x_2\rangle+a_{A^*}(\xi_3)\langle\xi_2,x_1\rangle
\end{eqnarray*}
and
\begin{eqnarray*}
(e_2,[e_1,e_3]_E)_-&=&-(\langle \xi_1\cdot_{A^*}\xi_2,x_3\rangle-\langle\xi_2,[x_1,x_3]_A\rangle+\langle\xi_1,x_3\cdot_{A}x_2\rangle\\
&&-\langle\xi_3,x_1\cdot_{A}x_2\rangle+\langle x_2,[\xi_1,\xi_3]_A\rangle-\langle \xi_3\cdot_{A^*}\xi_2,x_1\rangle)\\
&&+a_{A^*}(\xi_1)\langle\xi_2,x_3\rangle-a_{A^*}(\xi_3)\langle\xi_2,x_1\rangle-a_A(x_1)\langle x_2,\xi_3\rangle+a_A(x_3)\langle \xi_1,x_2\rangle,
\end{eqnarray*}
which implies  that condition (ii) in Definition \ref{MTLA} holds. \qed
\vspace{3mm}

{\bf Proof of Proposition \ref{pro:Manin-LSbi}}.
 Since the pairing $(\cdot,\cdot)_-$ is nondegenerate, $L_2$ is isomorphic to $L_1^*$,  the dual bundle of $L_1$, via $ \langle \xi,y\rangle=(\xi,y)_-$ for all $y\in\Gamma(L_1),~\xi\in\Gamma(L_2)$. Under this isomorphism, the skew-symmetric bilinear form $(\cdot,\cdot)_-$ on $E$ is given by
 $$(\xi+x,y+\eta)_-=\langle \xi,y\rangle-\langle y,\eta\rangle.$$
By Proposition $\ref{Dirac subbundles}$, both $L_1$ and $L_2$ are left-symmetric algebroids, their anchors are given by $a=\rho\mid_{L_1}$ and $a_*=\rho\mid_{L_2}$ respectively. We shall use $\delta,~~\dM$ and $\delta_*,~~\dM_*$ to denote their differential of left-symmetric algebroids and corresponding sub-adjacent Lie algebroids respectively.

By condition (ii) in Definition $\ref{MTLA}$, we deduce that the bracket between $x\in\Gamma(L_1)$ and $\xi\in\Gamma(L_2)$ is given by
\begin{eqnarray*}
x\star \xi&=&\huaL_x\xi+\frac{1}{2}\dM\langle x,\xi\rangle-R_{\xi}x+\frac{1}{2}\dM_*\langle x,\xi\rangle;\\
\xi\star x&=&\huaL_{\xi}x-\frac{1}{2}\dM_*\langle x,\xi\rangle-R_{x}\xi-\frac{1}{2}\dM\langle x,\xi\rangle.
\end{eqnarray*}
Thus, the multiplication $ \star  $ is given by $(\ref{operation})$.

It follows from Lemma $\ref{associator 1}$ that $I_1+I_2-I_3=0$. Since the anchor $\rho$ is a Lie algebroid homomorphism, we have $I_2=I_3=0$. Thus, $I_1=0$, i.e.$$\delta_*[x_1,x_2]_{L_1}-\frkL_{x_1}{\delta}_*{x_2}+\frkL_{x_2}{\delta}_*{x_1}=0,\quad \forall x_1,x_2\in\Gamma(L_1).$$
Similarly, we   have $$\delta[\xi_1,\xi_2]_{L_2}-\frkL_{\xi_1}{\delta}{\xi_2}+\frkL_{\xi_2}{\delta}{\xi_1}=0,\quad \forall~\xi_1,\xi_2\in\Gamma(L_2).$$ Thus,  $(L_1,L_2)$ is a left-symmetric bialgebroid.\qed
\vspace{3mm}

By Theorem  \ref{thm:equivalence2}, we have

\begin{cor}\label{cor:matched pair symplectic}
 Let $(A,A^*)$ be a left-symmetric bialgebroid. Then $(A^c,{A^*}^c)$ is a matched pair of Lie algebroids and the bracket on ${A}^c\oplus {A^*}^c$ is defined by
    \begin{eqnarray}\label{eq:mpbracket}
    [x+\xi,y+\eta]=[\xi,\eta]_{A^*}+L^*_x\xi-L^*_y\eta+L^*_\xi y-L^*_\eta x+[x,y]_A,\quad x,y\in\Gamma(A),\xi,\eta\in \Gamma(A^*).
    \end{eqnarray}
    Furthermore, $({A}^c\oplus {A^*}^c,\omega)$ is a symplectic Lie algebroid, where $\omega$ is given by \eqref{eq:defiomega}.
\end{cor}

By Proposition \ref{cor:hessianLA} and Proposition \ref{pro:LSbi-Manin}, we obtain
\begin{pro}
 Let $(M,\nabla,g)$ be a pseudo-Hessian manifold. Then $(TM\oplus T^*M,\star,{\id}+H^\sharp,(\cdot,\cdot)_-)$ is a pre-symplectic algebroid, where for all $e_1=x_1+\xi_1,e_2=x_2+\xi_2\in\Gamma(TM\oplus T^*M)$, the multiplication $\star $ is given by
$$
e_1\star e_2=(\nabla_{x_1} x_2+\huaL_{\xi_1}x_2-R_{\xi_2}x_1-\frac{1}{2}\dM_*(e_1,e_2)_+)+(\xi_1\cdot_{H} \xi_2+\huaL_{x_1}\xi_2-R_{x_2}\xi_1-\frac{1}{2}\dM(e_1,e_2)_+).
$$
Here $\huaL$ is given by \eqref{Lie der1}, $R$ is given by \eqref{eq:Rformula} and $(\cdot,\cdot)_+$ is given by \eqref{sym-form}.
\end{pro}

Assume that $(A,A^*)$ is a left-symmetric bialgebroid and $H\in \Gamma(A\otimes A)$. We denote by $G_H$ the graph of $H^\sharp$,  i.e. $G_H=\{H^\sharp(\xi)+\xi|~~\forall \xi\in A^*\}$.

\begin{thm} With the above notations,
$G_H$ is a Dirac structure of the pre-symplectic algebroid $(A\oplus
A^*,\star,\rho,(\cdot,\cdot)_-)$ given by Proposition
\ref{pro:LSbi-Manin}   if and only if  $H\in \Sym^2(A)$ and the following Maurer-Cartan type equation is satisfied:

\begin{equation}
\delta_{\ast}H-\llbracket H,H\rrbracket=0,
\end{equation}
where $\llbracket H,H\rrbracket$ is given by \eqref{brac2}.
\end{thm}
\pf
First it is easy to see that $G_H$ is isotropic if and only if $H\in \Sym^2(A)$.
By \eqref{operation}, we have
\begin{eqnarray*}
H^\sharp(\xi)\star\eta&=&\huaL_{H^\sharp(\xi)}\eta-\frac{1}{2}\dM\langle H^\sharp(\xi),\eta\rangle-R_{\eta}H^\sharp(\xi)-\frac{1}{2}\dM_{*}\langle H^\sharp(\xi),\eta\rangle;\\
\xi\star H^\sharp(\eta)&=&-R_{H^\sharp(\eta)}\xi-\frac{1}{2}\dM\langle H^\sharp(\xi),\eta\rangle+\huaL_{\xi}H^\sharp(\eta)-\frac{1}{2}\dM_{*}\langle H^\sharp(\xi),\eta\rangle,
\end{eqnarray*}
which implies that
$$ H^\sharp(\xi)\star\eta+\xi\star H^\sharp(\eta)=\xi\cdot_H\eta+\huaL_{\xi}H^\sharp(\eta)-R_{\eta}H^\sharp(\xi)-\dM_{*}\langle H^\sharp(\xi),\eta\rangle.$$
Then by $(\ref{homo})$, we have
\begin{eqnarray*}
&&(H^\sharp(\xi)+\xi)\star(H^\sharp(\eta)+\eta)\\
&=&H^\sharp(\xi)\cdot_{A}H^\sharp(\eta)+ H^\sharp(\xi)\star\eta+\xi\star H^\sharp(\eta)+\xi\cdot_{A^*}\eta\\
&=&H^\sharp(\xi\cdot_H\eta)+\huaL_{\xi}H^\sharp(\eta)-R_{\eta}H^\sharp(\xi)-\dM_{*}\langle H^\sharp(\xi),\eta\rangle-\llbracket H,H\rrbracket(\xi,\cdot,\eta)+\xi\cdot_H\eta+\xi\cdot_{A^*}\eta.
\end{eqnarray*}
Thus, $G_H$ is integrable if and only if for all $\xi,\eta\in\Gamma(A^*)$,
\begin{equation}\label{integ}
H^\sharp(\xi\cdot_{A^*}\eta)=\huaL_{\xi}H^\sharp(\eta)-R_{\eta}H^\sharp(\xi)-\dM_{*}\langle H^\sharp(\xi),\eta\rangle-\llbracket H,H\rrbracket(\xi,\cdot,\eta).
\end{equation}
On the other hand, we have
\begin{eqnarray}
\nonumber\delta_{*}H(\zeta,\xi,\eta)&=&a_A(\zeta)\langle H^\sharp(\xi),\eta\rangle-a_A(\xi)\langle H^\sharp(\zeta),\eta\rangle\\
\nonumber&&-\langle H^\sharp(\xi),\zeta\cdot_{A^*}\eta\rangle+\langle H^\sharp(\zeta),\xi\cdot_{A^*}\eta\rangle-\langle[\zeta,\xi]_{A^*},H^\sharp(\eta)\rangle\\
\label{eq:fff}&=&\langle H^\sharp(\xi\cdot_{A^*}\eta)-\huaL_{\xi}H^\sharp(\eta)+R_{\eta}H^\sharp(\xi)+\dM_{*}\langle H^\sharp(\xi),\eta\rangle,\zeta\rangle.
\end{eqnarray}
By $(\ref{integ})$ and \eqref{eq:fff}, $G_H$ is a Dirac structure if and only if
$$\delta_{\ast}H(\zeta,\xi,\eta)-\llbracket H,H\rrbracket(\zeta,\xi,\eta)=0.$$
The proof is finished.\qed

\end{document}